\documentclass[fleqn,twoside]{article}
\usepackage[headings]{espcrc2}

\readRCS
$Id: espcrc2.tex,v 1.2 2004/02/24 11:22:11 spepping Exp $
\usepackage{graphicx}
\usepackage[figuresright]{rotating}

\title{Heavy meson semileptonic decays in two dimensions in the large $N_c$}

\author{Jorge Mondejar\address[ECM]{Dept. d'Estructura i Constituents de la Materia, U. Barcelona, \\
	Diagonal, 647, E-08028 Barcelona, Spain} }

\begin{document}

\begin{abstract}
We study QCD in $1+1$ dimensions in the large $N_c$ limit using light-front Hamiltonian perturbation theory in the $1/N_c$ expansion. We use this formalism to exactly compute hadronic transition matrix elements for arbitrary currents at leading order in $1/N_c$, which we use to write the semileptonic differential decay rate of a heavy meson and its moments. We then compare with the results obtained using an effective field theory approach based on perturbative factorization, with the intention of better understanding quark-hadron duality. A very good numerical agreement is obtained between the exact result and the result using effective theories.
\vspace{1pc}
\end{abstract}

\maketitle

\section{Introduction}

The results presented here are part of a work done in collaboration with Antonio Pineda and Joan Rojo \cite{Mondejar_et_al:2006}.

Asymptotic freedom can be seen as the first example of 
factorization between high and low energies, since 
it dictates that Green functions 
at high Euclidean energies ($Q^2$) can be described by 
perturbation theory up to corrections suppressed by 
powers of $\Lambda_{QCD}$ over $Q$. 
Therefore, the use of the operator product expansion (OPE) in processes where the
relevant momentum scale is large and Euclidean is safe. This is quite 
restrictive, since, in most of the cases, it can only be tested with experiment
through dispersion relations, which involve measurements up to
arbitrarily high energies. What one usually does is to try to directly apply the 
same perturbative factorization techniques to observables living in the
Minkowski regime. In practice this means to perform the analytic continuation of 
approximate perturbative results obtained in the Euclidean region to the 
Minkowski region, but such calculations do not come from first 
principles. This problem affects the OPE and effective field
theories that are built using perturbative factorization techniques 
aiming to factorize high from low energies, and it is
usually stated as duality violations. We will follow here the definition of
\cite{Shifman:2000jv} for duality violations. 

One can quantify the discrepancy between the exact result and the one 
using perturbative factorization in the large $N_c$ limit of QCD \cite{hooft1}. In 
this case one finds a clear discrepancy between both results 
in the physical cut of the Green functions, where one has infinitely 
narrow resonances on the one hand 
and an smooth function on the other. This can be further quantified in the 
't Hooft model \cite{hooft2}, which we will consider in what follows. 

The specific observable we use to illustrate this discussion is the differential semileptonic inclusive decay of a heavy meson: $H_Q\to X l \nu$. The duality violations in this case are maximal, but if we consider the Mellin moments of the differential decay rate we find that there is a very good agreement between the exact and the perturbative result.

In sec. 2 we analyze QCD$_{1+1}$ in the light front. In sec. 3 we compute the hadronic differential decay rate and its moments. In sec. 4 we construct an effective theory to compute these quantities at one loop, and compare the two results. Finally, in section 5 we present our conclusions.

\section{QCD$_{1+1}$ in the light front}

The QCD lagrangian is given by
\begin{equation}
\mathcal{L}=-\frac{1}{4}G^a_{\mu\nu}G^{a,{\mu\nu}}+
\sum_i \bar{\psi}_i\left( i\gamma^{\mu} D_{\mu}-m_i +i\epsilon \right) \psi_i
\,,
\end{equation}
where $D_{\mu}=\partial_{\mu}+igA_{\mu}$ and $i$ labels the flavor. 

We work in light-cone coordinates: we define two light-like vectors,
\begin{equation}
n_-^{\mu}=(1,1)\ , \quad n_+^{\mu}=(1,-1) \  ,
\end{equation}
and define the light-cone coordinates as
\begin{equation}
x^+\equiv n_+\cdot x=( x^0+x^1), 
\quad
x^-\equiv n_-\cdot x=( x^0-x^1) \ . 
\end{equation}
In this coordinates the usual quantization gauge is
$A^{+}\equiv n_+\cdot A=0$, the so called light-cone gauge. As for the mass fields $\psi_i$, we split them in
\begin{equation}
\psi_+=\frac{1}{4}\gamma^-\gamma^+\psi \ , \quad \psi_-=\frac{1}{4}\gamma^+\gamma^-\psi \ .
\end{equation}
With these definitions, the QCD lagrangian looks like
\begin{equation}
\mathcal{L}=	\frac{1}{8}(\partial^+A^{-})^2+\sum_i\left(\psi_{i+}^{\dagger}(i\partial_-+gA^-)\psi_{i+} \right.
\end{equation}
\begin{displaymath}
\left. +\psi_{i-}i\partial^+\psi_{i-}-m_i(\psi_{i+}^{\dagger}\gamma^0\psi_{i-}+\psi_{i-}^{\dagger}\gamma^0\psi_{i+})\right) \ .
\end{displaymath}
As our quantization frame, we choose to quantize at $x^+=$ constant, which means that $x^+$ plays the role of time, and $x^-$ that of space in our equations.
In this quantization frame, neither $A^-$ nor $\psi_{i-}$ are dynamical fields, so we can integrate them out and construct the Hamiltonian
\begin{eqnarray}
P^-&=&\sum_i\int dx^-dy^- \\
& &\times\{ \frac{-i m_i^2}{4} \psi^{\dagger}_{i+}(x^-)
\epsilon(x^--y^-)\psi_{i+}(y^-)  \nonumber
\end{eqnarray}
\begin{displaymath}
-\sum_{j}\frac{g^2}{4}
\psi^{\dagger}_{i+}t^a\psi_{i+}(x^-)|x^--y^-|
\psi^{\dagger}_{j+}t^a
\psi_{j+}(y^-)\}
\  , 
\end{displaymath}
where
\begin{equation}
\epsilon(x)=
\left\{
\begin{array}{ll}
-1\ , & x<0 \,,\\
0 \ , & x=0 \,,\\
1 \ , & x>0 \,.
\end{array}\right.
\\
\end{equation}
By solving the eigenstate equation $P^-|n\rangle=P_n^-| n \rangle$ one obtains the basis of states on which the Hilbert space of physical states can be spanned. Here we will focus on the meson sector of this space, and we will generically label the state as $|ij;n\rangle$, 
where $i$ labels the flavor of the valence quark, $j$ labels the 
flavor of the valence antiquark and $n$ labels the excitation 
of the bound state.  
The solution to the eigenstate equation can be obtained from the large $N_c$ limit solutions 
 within a systematic 
expansion in $1/N_c$ using standard time-independent quantum perturbation 
theory. Up to $\mathcal{O}(1/N_c)$, it has the following structure
 \begin{equation}
|ij;n\rangle=|ij;n\rangle^{(0)}
+
\sum_{m,n'}\sum_k|ik;n'\rangle^{(0)}|kj;m\rangle^{(0)}
\end{equation}
\begin{displaymath}
\times{}^{(0)}\langle ik; n'|{}^{(0)}\langle kj;m|P^-|ij;n\rangle^{(0)}
\frac{1}{P_n^{(0)-}-P_m^{(0)-}-P_{n'}^{(0)-}}
\,,
\end{displaymath}
where the second term in the expression is $1/\sqrt{N_c}$ suppressed: in our calculations we are staying at leading order in $1/N_c$, but we have to keep this term because it gets enhanced by $\sqrt{N_c}$ when computing transition matrix elements. $|ij;n\rangle^{(0)}$
represents the eigenstate solution in the large $N_c$
limit,
\begin{eqnarray}
|ij;n\rangle^{(0)}&=&\frac{1}{\sqrt{N_c}}
\int_0^{P_n^+} \frac{dp^+}{\sqrt{2(2\pi)}}\phi^{ij}_{n}\left(\frac{p^+}{P_n^+}\right) \\
& &\times a_{i,\alpha}^{\dagger}(p)
b_{j,\alpha}^{\dagger}(P_{n}-p)|0\rangle
\,, \nonumber
\end{eqnarray}
where $\alpha$ is the color index, $\phi_{n}^{ij}$ is the solution to the 't Hooft equation,
 and the state is normalized as
\begin{eqnarray}
{}^{(0)}\langle ij;m|i'j';n \rangle^{(0)}&=&2\pi 2P_n^{(0)+}\delta_{mn}\delta_{ii'}
\delta_{jj'} \\
& &\times \delta(P_m^{(0)+}-P_n^{(0)+})
\,. \nonumber
\end{eqnarray}

With this we can compute transition matrix elements, $\langle c s ;m|{\bar \psi}_{c}\Gamma Q|Q s;n\rangle$. For reasons of space, we cannot show the formulas here, and refer the reader to \cite{Mondejar_et_al:2006}.

\section{Semileptonic differential decay rate: hadronic computation}

We consider the semileptonic heavy meson decay: $H_Q \rightarrow X_c l_a \bar l_b$, 
where $H_Q$ represents a bound state made of a heavy quark $Q$ and a light (spectator) 
quark $s$, $X_c$ represents any hadronic final state with $c$ (hard-collinear) 
flavour content and $l_{a,b}$ represent massless leptons. We will consider the situation on which 
the spectator, $\psi_{s}$, and hard-collinear, $\psi_c$, quarks have different flavour in order to avoid annihilation and Pauli 
interference terms. 
This decay has already been studied in the past, we will follow here the work 
of Bigi et al. \cite{Bigi:1998kc}. The authors considered the flavour 
changing weak interaction 
\begin{equation}
\mathcal{ L}_{\rm weak}^V=-\frac{G}{\sqrt{2}}\bar \psi_c \gamma_{\mu}Q \bar
l_a \gamma^{\mu} l_b
\,.
\end{equation}
The total decay width can be written 
as
\begin{equation}
\Gamma_{H_Q}=\frac{G^2}{M_{H_Q}}\int \frac{d^2q}{(2\pi)^2} 
Im \Pi_{\mu\nu}(q)
Im T^{\mu\nu}(q)
\,,
\end{equation}
where $\Pi_{\mu\nu}(x)$ and $ T_{\mu\nu}(x)$ are defined as
\begin{equation}
  \Pi_{\mu\nu}(x)= i\,\langle 0|T \left \{ \bar{l}_a(x) \gamma_\mu l_b(x)
\, \bar{l}_b(0) \gamma_\nu l_a(0)\right\}|0\rangle
\end{equation}
\begin{displaymath}
 T^{\mu\nu}(x)= i\,\langle H_Q |T \left \{ \,
\bar{Q}(x) \gamma^\mu \psi_c(x) \,\bar{\psi}_c(0) \gamma^\nu Q(0)\right\}|H_Q\rangle ,
\end{displaymath}
and their Fourier transform as 
\begin{eqnarray}
\Pi_{\mu\nu} (q)&=&\int {\rm d}^2 x \, e^{iqx} \Pi_{\mu\nu}(x)
\\
T^{\mu\nu}(q)&=&\int d^2x e^{-iqx}T^{\mu\nu}(x)
\,. \nonumber 
\end{eqnarray}
The leptonic tensor can be easily calculated; to calculate the hadronic tensor we just have to insert a complete set of intermediate states: in the large $N_c$ limit this set will consist exclusively of mesons, so we can use the transition matrix elements we found previously.  In the end, the differential decay rate reads (the details can be found in \cite{Mondejar_et_al:2006})
\begin{equation}
\label{dxh}
\frac{d\Gamma^{hadr}}{dx}=\sum_{M_{n}\le M_{H_Q}} \frac{G^2}{4\pi}\frac{M_{H_Q}^2-M_n^2}{M_{H_Q}}
\end{equation}
\begin{displaymath}
\times\left[\int_0^1dz \phi^{cs}_n(z)\phi_{H_Q}(z)\right]^2
 \delta\left(x-1+\frac{M_n^2}{M_{H_Q}^2} \right), \nonumber
\end{displaymath}
where
\begin{equation}
x\equiv \frac{q^+}{P_{H_Q}^+} \ .
\end{equation}
The moments are defined as
\begin{equation}
M_N \equiv \int_0^1 dx x^{N-1} \frac{d\Gamma}{dx}
\,.
\end{equation}
We can insert (\ref{dxh}) in this expression right away, but  since our interest here is to compare the result with that of an effective theory, we will give an alternative expression for the moments which is valid up to $N\sim m_Q/\beta$ (again, the details of the calculation, and the approximations made, can be found in \cite{Mondejar_et_al:2006}),
\begin{eqnarray}
\label{hadrmom}
M_N^{hadr}&\simeq &
\frac{G^2M_{H_Q}}{4\pi}\frac{m_{Q,R}^2}{M_{H_Q}^2}
\left( 1-\frac{m_{c,R}^2}{m_{Q,R}^2}\right)^N 
\\
& &\times\int_0^1 
 \frac{dx}{x^2} x^{N}\phi_{H_Q}^2(x) \ , \nonumber
\end{eqnarray}
where $m_{x,R}^2=m_x^2-\beta^2$.
 This formula has a precision of $\mathcal{O}(\beta^2/m_Q^2)$ when $N\sim 1$ which reduces to $\mathcal{O}(\beta /m_Q)$ when $N\sim m_Q/\beta$.

\section{Effective theory}

\begin{figure}[h]
\begin{center}
\includegraphics*[width=1.0\columnwidth]{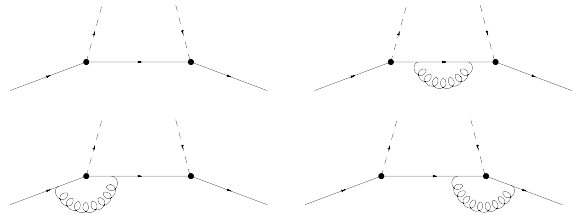}
\caption{Diagrams involved in the construction of the effective theory at one loop. The dotted lines represent the outgoing leptons.} 
\label{g1}
\end{center}
\end{figure}

To construct the effective theory we switch to a partonic picture, in which the decay of the initial meson becomes just the decay of the heavy quark into a light quark. The decay is given by the imaginary part of the diagrams shown in Figure \ref{g1}. 

We work in the kinematics in which the momentum carried away by the leptons is
\begin{equation}
q^+=x p^+_Q \ , \quad q^-=0 \ .
\end{equation}
This means that the light outgoing quark resulting from the decay has the following momentum:
\begin{equation}
p_c^+=p^+_Q(1-x) \ , \quad p_c^-=p_Q^- \ .
\end{equation}
The important point is that $p_c^-=p_Q^-=m_Q$, which in our light front quantization frame means that the propagating light quark in the diagrams has a very large ``energy", and that the process happens at very short ``times". So, we can safely integrate out the light quark from our lagrangian. The resulting lagrangian will be essentially that of HQET, plus a local vertex in $x^+$, the imaginary part of which will represent the decay process. The construction of the effective theory amounts then to computing in full QCD the diagrams in Figure \ref{g1}, and then matching to an effective vertex in which the intermediate light quark will have disappeared. The effective vertex is
\begin{eqnarray}
\label{LI}
\mathcal{L}_I&=&
-\frac{G^2}{2\pi}(\partial^+\phi)\left( \frac{m_{Q,R}}{i\partial^+}Q_{+}\right)^{\dagger}\frac{1}{i\partial^+-\frac{m_{c,R}^2-i\epsilon}{m^2_{Q,R}}
i\partial^+}\nonumber\\
& &\times\left( 
\frac{m_{Q,R}}{i\partial^+}Q_{+}\right)(\partial^+\phi^{\dagger}) 
\, ,
\end{eqnarray}
where $\phi$ is a pseduoscalar massless field that represents the (massless) outgoing leptons.
The differential decay rate is then given by
\begin{equation}
{d \Gamma \over dx}^{pert} = \frac{1}{M_{H_Q}}\frac{1}{2(2\pi)x} \frac{G^2}{2\pi} 
(M_{H_Q}x)^2 2 {\rm Im} T_{eff} 
\end{equation}
\begin{displaymath}
= 
\frac{G^2M_{H_Q}}{4\pi}\frac{m_{Q,R}^2-m_{c,R}^2}{m^2_{Q,R}}\left(
\frac{m^2_{Q,R}}{M^2_{H_Q}}\right)  \frac{1}{x}\phi_{H_Q}^2\left( \frac{x}{1-\frac{m_{c,R}^2}{m_{Q,R}^2}}\right)
\,.
\end{displaymath}
Comparing with (\ref{dxh}) we see the maximal duality violation: one result is a sum of deltas whereas the other is a smooth function. However, if we compute the moments, we find
\begin{eqnarray}
\label{MNscet}
M_N^{pert}&=&\frac{G^2M_{H_Q}}{4\pi}\frac{m_{Q,R}^2}{M_{H_Q}^2}\left(1-\frac{m_{c,R}^2}{m_{Q,R}^2}\right)^N  \\
& &\times
\int_0^1 dx x^{N-2}\phi_{H_Q}^2(x)
\, , \nonumber
\end{eqnarray}
the same expression as (\ref{hadrmom}), with the same precision. Thus, up to the precision we are working with, there are no duality violations for the moments. Actually, a numerical analysis (see \cite{Mondejar_et_al:2006}) shows that the numerical agreement is very good up to high moments. In  \cite{Bigi:1998kc} it is shown for the inclusive decay width that differences appear at $\mathcal{O}(1/m_Q^9)$ (they are strongly suppressed, but they are there). In \cite{Mondejar_et_al:2006} it is also shown that, although very good results are found for the moments, things don't look so well if we average over smaller ranges. Namely, for
\begin{equation}
\label{deltax}
\int_{x_n-\delta x}^{x_n+\delta x}\frac{d \Gamma}{dx} dx
\,,
\end{equation}
where $x_n$ is any $x$ which satisfies the delta in (\ref{dxh}), we find that, first, we have to fine-tune the value of $\delta x$ for the hadronic and perturbative results to agree at leading order, but even then,  they differ at $\mathcal{O}(\beta^2/m_Q^2)$.

\section{Conclusions}

To sum up, we have computed the differential semileptonic inclusive decay of a heavy meson and its moments, through both a hadronic calculation and a perturbative calculation (using an effective theory), and found that, whereas a comparison between the two results is impossible for the differential decay rate, the moments show no duality violations with the precision we have worked with; however, other observables, like the one defined in (\ref{deltax}) don't work so well. So, in conclusion, effective theories can only be a good approximation for inclusive observables on which one averages over a large fraction of the final bound states. They are not suited for point-to-point comparisons, or comparisons between arbitrarily smeared functions.

\end{document}